\documentclass[manuscript]{aastex631}

\newcommand{\footurlCast}{http://www.e-callisto.org}

\newcommand{\footurlGit}{https://github.com/desertfireballnetwork/freeture\_DFN}

\newcommand{\footurlRS}{https://manual.raspberryshake.org/specifications.html}

\newcommand{\footurlRSv}{https://manual.raspberryshake.org/\_downloads/SpecificationsforRaspberryShakeV4.pdf}

\usepackage{amsmath,amsbsy}
\usepackage{colortbl}

\graphicspath{{./}}

\begin{document}
\title{The Scientific Observation Campaign of the Hayabusa-2 Capsule Re-entry}

\author{E. K. Sansom}\affiliation{Space Science and Technology Centre, Curtin University, Australia}

\author{H. A. R. Devillepoix}\affiliation{Space Science and Technology Centre, Curtin University, Australia}

\author{M. -Y. Yamamoto}\affiliation{School of Systems Engineering, Kochi University of Technology, Japan}

\author{S. Abe}\affiliation{Nihon University, Japan}

\author{S. Nozawa}\affiliation{Ibaraki University, Japan}

\author{M. C. Towner}\affiliation{Space Science and Technology Centre, Curtin University, Australia}

\author{M. Cupák}\affiliation{Space Science and Technology Centre, Curtin University, Australia}

\author{Y. Hiramatsu}\affiliation{School of Geosciences and Civil Engineering, College of Science and Engineering, Kanazawa University, Japan}

\author{T. Kawamura}\affiliation{Institut de Physique du Globe de Paris, France}

\author{K. Fujita}\affiliation{Institute of Space and Astronautical Science, Japan Aerospace Exploration Agency, Japan}

\author{M. Yoshikawa}\affiliation{Institute of Space and Astronautical Science, Japan Aerospace Exploration Agency, Japan}

\author{Y. Ishihara}\affiliation{JAXA Space Exploration Center, Japan Aerospace Exploration Agency, Japan}

\author{I. Hamama}\affiliation{Graduate School of Engineering, Kochi University of Technology, Japan; National Research Institute of Astronomy and Geophysics, Helwan 11421, Cairo, Egypt}

\author{N. Segawa}\affiliation{Faculty of information Science and Engineering, Kyoto Sangyo University, Japan} %

\author{Y. Kakinami}\affiliation{Space Information Center, Hokkaido Information University, Japan}

\author{M. Furumoto}
\author{H. Katao}\affiliation{Research Center for Earthquake Prediction, Disaster Prevention Research Institute, Kyoto University, Japan}

\author{Y. Inoue}\affiliation{Graduate School of Engineering, Kochi University of Technology, Japan}

\author{A. Cool}\affiliation{The River Murray International Dark Sky Reserve, South Australia}

\author{G. Bonning}
\affiliation{Research School of Earth Sciences, The Australian National University, Australia}

\author{R. M. Howie}\affiliation{Space Science and Technology Centre, Curtin University, Australia}

\author{P. A. Bland}\affiliation{Space Science and Technology Centre, Curtin University, Australia}

\begin{abstract}
On 5th December 2020 at 17:28 UTC, the Japan Aerospace Exploration Agency's Hayabusa-2 sample return capsule came back to the Earth. It re-entered the atmosphere over South Australia, visible for 53 seconds as a fireball from near the Northern Territory border toward Woomera where it landed in the the Woomera military test range. A scientific observation campaign was planned to observe the optical, seismo-acoustic, radio and high energy particle phenomena associated with the entry of an interplanetary object. A multi-institutional collaboration between Australian and Japanese universities resulted in the deployment of 49 instruments, with a further 13 permanent observation sites. The campaign successfully recorded optical, seismo-acoustic and spectral data for this event which will allow an in depth analysis of the effects produced by interplanetary objects impacting the Earth's atmosphere. This will allow future comparison and insights to be made with natural meteoroid objects.

\end{abstract}

\section{INTRODUCTION }
\label{sec:intro}
When interplanetary material intersects the Earth, the atmospheric resistance causes intense heating. These objects can be seen as meteors in the night sky, or as fireballs if they are especially bright. Natural objects such as meteoroids will usually fragment and vaporise, though some may reach the ground as meteorites.  

The understanding of the physical processes that occur during the atmospheric entry, from before the visible meteor phenomena begins, to the free fall stage after ablation ceases, is still mostly theoretical. Rare observations have been obtained post-hoc, for example where meteorites have been recovered and recorded infrasound and seismic data has then been investigated \citep{brown_analysis_2008}, or occasional serendipitous ionosonde measurements, as reviewed by \citep{kereszturi_review_2021}. However, larger events that typically come from asteroidal sources are rare and unpredictable. In order to study them, large spatial and temporal coverage is required. This severely restricts the ability to observe using instruments with a narrow field of view, or record faint phenomena associated with a fireball that would require a dense network of sensors.  

The planned re-entry of an interplanetary spacecraft is a unique opportunity to test sensors and record aspects of fireball phenomena that are impossible to collect for sporadic, natural events. The first such opportunity was the return of the Stardust mission in 2006, where a single 4-station infrasound array was deployed \citep{Revelle2007, Edwards2007}. The return of the Hayabusa-1 spacecraft (called Hayabusa at that time) and sample return capsule (SRC) in 2010 was the first another such opportunity. It landed on June 13, 2010 in rural South Australia within the Woomera military test range. The ground based observations were primarily set up to aid in the recovery of the SRC \citep{fujita2011overview}. Though trajectory-based data were also sought in order to ascertain the environment the SRC was exposed to (temperatures, pressures).  

At that time, several of the authors tried to detect the shock waves coming from the Mach cone of the SRC by using 5 infrasound sensors of Chaparral Physics Model-25 and Model-2 on ground. The Hayabusa-1 re-entry was slightly different from the recent Hayabusa-2 re-entry because the mother spacecraft (S/C) of Hayabusa-1 itself suffered a partial malfunction and also re-entered the atmosphere with the SRC (though did not land). Therefore, multiple shock waves induced by both the SRC and multiple fragmented parts of the S/C were clearly observed by all 5 sensors. At the same time, over 20 seismometers simultaneously detected N-type signals with an air-to-ground coupling process (\citealt{yamamoto2011detection, ishihara2012infrasound}).  

On the 5th of December 2020, JAXA's Hayabusa-2 SRC returned to the Earth after collecting multiple samples from asteroid Ryugu. 
Once again, this re-entry was another opportunity to observe a pre-determined fireball event. Hayabusa-2 was planned to land in Australia on 5th Dec. 2020, in the same desert area as the Hayabusa-1 SRC. To observe this, a specific science observation campaign was planned between Japanese and Australian institutions, separate to the JAXA mission and engineering teams who focused on the tracking and recovery of the SRC.
The goal of this science observation campaign was to observe various phenomena from the high altitude entry to the end of the luminous trajectory. In particular, the campaign focused on acquiring data to characterise the shock wave produced by the hyper-sonic re-entry, for which the flight path of the SRC through the atmosphere must be well known.  

The Desert Fireball Network (DFN) in Australia observes 2.5 million km$^2$ of skies, and is designed to triangulate fireball phenomena to recover meteorites with orbits \citep{Howie2017Howbuildcontinental}. DFN cameras are deployed in South Australia, covering the Hayabusa-2 return site, and have been operational in this area for several years. Observing the SRC re-entry using these instruments provides a known, high precision flight path for the reduction of non-optical data. This event also provides the DFN science team with an opportunity to validate trajectory models and orbital calculation methods used for natural bodies by comparing results with the known trajectory and orbit of Hayabusa-2.  

To maximise this opportunity, a variety of instruments were temporarily deployed to cover a wide range of observations, both optical and non-optical. These included additional DFN systems, seismo-acoustic sensors for shock wave characterisation, UHF antenna and high energy particle detectors. This summary paper will describe the instruments deployed and the preliminary campaign results. 

\section{Instrumentation}\label{sec:instrumentation}

Here we describe the instrumentation orientated towards scientific observations of the Hayabusa-2 SRC re-entry. 49 instruments were deployed along the planned re-entry trajectory, including optical, seismo-acoustic, radio and high energy particle detectors (Table \ref{tab:instruments}). These were augmented by a further 13 permanent Desert Fireball Network sites, each capturing both all-sky video and long exposure still images.  

\begin{table*}
\caption{Summary of instruments deployed for scientific observations of the Hayabusa-2 SRC trajectory. Where ($^+$) indicates additional instruments available at permanent Desert Fireball Network sites within range. }\label{tab:instruments}
\centering
\setlength{\tabcolsep}{10pt} 
\begin{tabular*}{\textwidth}{@{}l | l | l | l | l}
\hline \hline
 \textbf{Instrument} & 
    \textbf{Functions} & 
    \textbf{Field of View} & 
    \textbf{Lead} &
    \textbf{\# sensors}\\
    &&&\textbf{team}&\textbf{deployed}\\
\hline
Long exposure
        & $\circ$ trajectory triangulation 
        & all-sky 
        & CU 
        & 4 ($^+$13)
        \\ still image &&&&\\
        \hline 
    Fixed video & 
        $\circ$ trajectory triangulation (testing) 
        & all-sky 
        & CU 
        & 5 ($^+$13)
        \\& $\circ$ light curves&&&\\
        \hline
    Tracking video 
        & $\circ$ spectroscopy of fireball and train 
        & 74 deg 
        & NU 
        & 1\\
        \hline
    Infrasound 
        & $\circ$ record audible sound and infrasound 
        & omnidirectional 
        & KUT 
        & 28
        \\
        \hline 
    RS 1D 
        & $\circ$ record 1D seismic 
        & Vertical
        & KUT 
        & 2\\
        \hline 
    RS 3D 
        & $\circ$ record 3D seismic 
        & Vertical/East/North
        & KUT 
        & 2\\
        \hline
    RS\&B 
        & $\circ$ record 1D seismic
        & Vertical
        & CU
        & 3
        \\ & $\circ$ record audible sound and infrasound 
        & omnidirectional &&\\
        \hline
    UHF antenna 
        & $\circ$ passive detection of UHF radio waves 
        & Zenith \&
        & IU
        & 2
        \\ && parallel to trajectory&\\
        \hline 
    Energetic particle 
        & $\circ$ detection of any possible ionising
        & omnidirectional 
        & CU 
        & 2
        \\detector &radiation&&&\\[2pt]
\hline \hline
\end{tabular*}
\end{table*}

\subsection{Optical instruments}\label{sec:optic_instr}

\subsubsection{Long-exposure still images}\label{sec:long_exposure_instruments}
 
The standard Desert Fireball Network observatory takes long-exposure still images. Each consists of a Nikon D810 DSLR, with a Samyang 8 mm fisheye lens, capturing an all-sky image with approximately 1-2 arcmin/pixel spatial precision \citep{Howie2017Howbuildcontinental}. Each long-exposure is typically 27 sec. long, captured every 30 sec. (resulting in 3 sec. down time), and a liquid crystal shutter encodes timing information for any moving objects to allow precise timing \citep{Howie2017Submillisecondfireballtiming, Howie2019Absolutetimeencoding}. The DFN typically operates with all observatories taking pictures at 00 and 30 seconds past the minute, in a coordinated manner. In the case of Hayabusa-2 campaign, this would have resulted in a coordinated dead time, where no camera was observing. Consequently, selected cameras on the network were reconfigured to observe at 20 and 40 seconds past the minute, to allow overlapping observations.  

These systems are the same as the ones used by the DFN to recover several meteorites \citep{sansom2020murrili, Devillepoix2018DingleDellmeteorite}. The brightness of natural fireball phenomenon usually targeted by the DFN has a limiting magnitude of $\sim$0 Mag. The Hayabusa-2 SRC was predicted to be –5 Mag at its peak brightness. The amount of the SRC trajectory that could be recorded was predicted to be incomplete using the standard DFN systems, and a more sensitive video system was used to ensure continuous capture of the fireball.   

\subsubsection{All-sky fixed video}\label{sec:vid_data}
The latest generation of DFN observatories, known as \textit{DFNEXT}, were introduced in 2017 and now make up most of the Global Fireball Observatory outside Australia \citep{2020P&SS..19105036D}. These are equipped with a digital video cameras (Pont Grey/FLIR BFLY-U3-23S6M-C) with Fujinon fisheye F/1.4, 1.8 mm lenses with approximately 11 arcmin/pixel resolution. This addition in parallel to the still high-resolution imager was introduced to not only observe fainter meteors than the still photographs, but also provide observational coverage during the dead time between long exposures (3\,s out of 30\,s is 10\% loss), yield better photometry, and allow observation during the daytime. Although the Australian observatories are mostly using the older \textit{DFNSMALL} system, an effort was made for the Hayabusa-2 campaign to mostly use these newer \textit{DFNEXT} systems in order to capture digital video records.  

The control software for these digital video cameras is a modified version of \textit{Freeture} \citep{colas2020fripon} to detect and save fireball images (code available on 
Github\footnote{\footurlGit , commit hash \textit{be0623031655f6ff71cac44c3ba3b62e25ed9c17}} was used for the campaign). This software typically assesses frames for fireballs, and only saves positive detections. For Hayabusa-2, a special mode was added to the software, in which all video frames are recorded for a set period of time to circumvent potential issues with the detection software. Frames are saved as 8 bit lossless compressed FITS files, 30 frames per second, with gain at 29. In order to capture fainter stars for precise astrometric calibration, in the hours leading up to the event before the Moon rose, 4 second long exposures were captured every 10 minutes. 

\subsubsection{Narrow angle video}\label{sec:narrow_ang_vid}

At Marla --the most up-range temporary site-- a narrow angle video setup was deployed, aiming to catch the earliest hint of light produced by the fireball. A Fujinon 16\,mm, f/1.4, aimed at the 90\,km predicted altitude point, with the long axis of the $1920\times1200$ sensor (Point Grey BFLY-U3-23S6M-C) oriented along the trajectory. Although this set-up should have had in theory a sensitivity of $+5$ stellar magnitudes deeper than the all-sky video setup (Sec. \ref{sec:vid_data}), and a 73 arcsec/pixel
resolution, a focus issue during instrument deployment prevented from reaching this value. Nevertheless, the accurate pre-pointing achieved by the operators (TW and GB) enabled the successful capture of what is the earliest astrometry and photometry data for this fireball. No long-exposure calibration image was collected for this system, however enough stars are detectable in a single video frame to provide astrometric reference points.

\subsubsection{Spectral video}
\begin{figure}[h!]
    \centering
    \includegraphics[width = 0.5\columnwidth, angle=270]{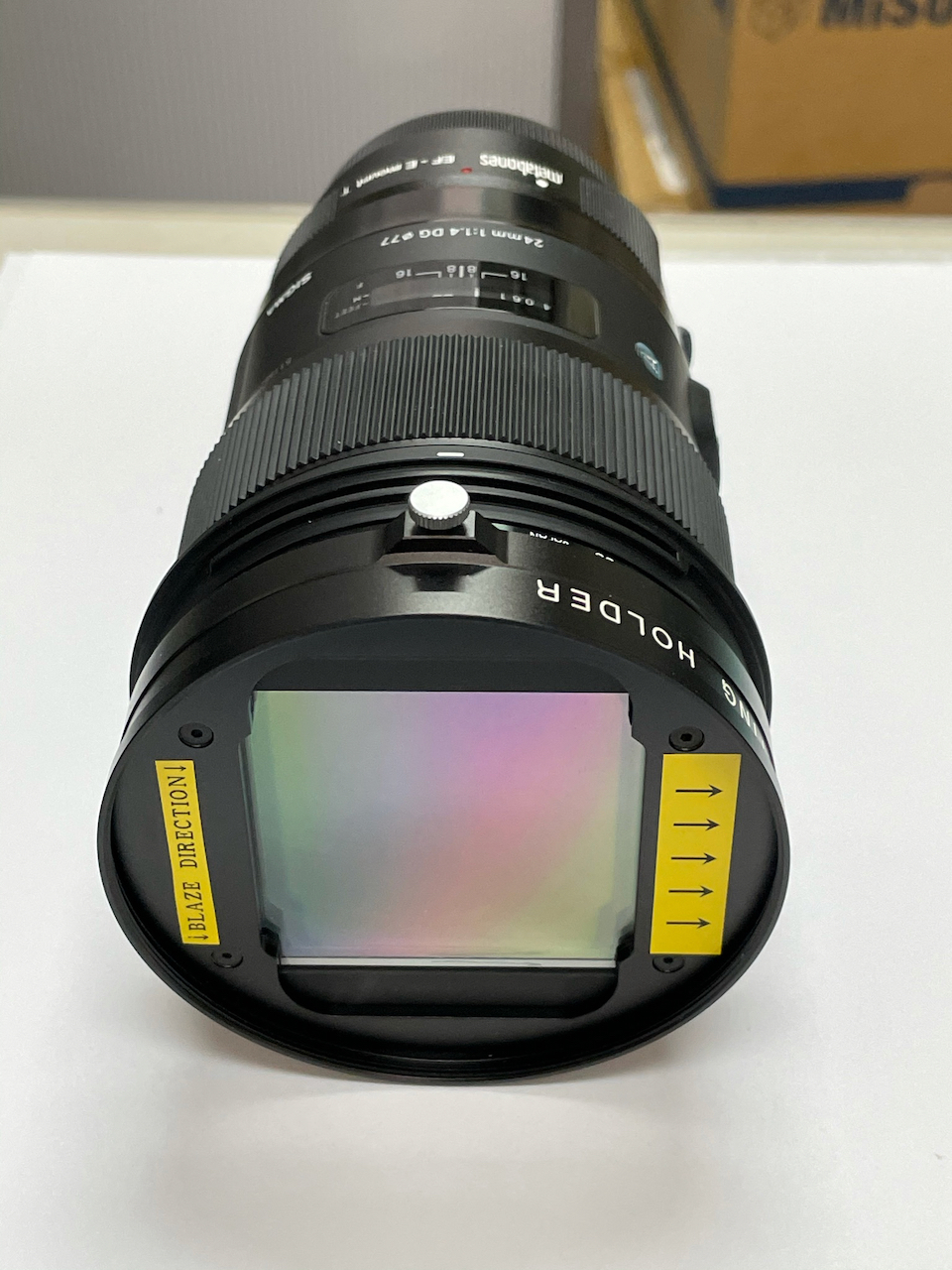}
    \caption{24\,mm lens and 600 grooves/mm grating used for recording the spectrum of the Hayabusa-2 SRC fireball.
    The spectral dispersion direction is shown by the arrows.}
    \label{fig:lens_grating}
\end{figure}

Video rate spectroscopic observations of meteors provide valuable information about emission processes in the atmosphere. Recently, high sensitive large format colour cameras, such as Sony A7S (ILCE-7S), have become popular due to their affordability. The main advantage of a large format sensor is a higher spatial resolution in digital format, which enables highly accurate analysis for spectroscopy. A de-Bayered Sony A7sII (ILCE-7SM2, 12 Mpx) camera with a Sigma 24\, mm f/1.4 lens was fitted with a 600 grooves/mm transmission grating (as seen in \ref{fig:lens_grating}). This monochromatic system is an optimal set up with the high sensitivity and linear response, enabling reliable yet straightforward analysis.

A Nikon 14-24\,mm f/2.8 (operated at 14\,mm) lens with a 300\,g/mm grating in front of the lens was also set up on a FLI KL4040 camera in video mode, though unfortunately this system failed to capture data because of a software failure.

\newpage
\subsection{Seismo-Acoustic instruments}
\begin{figure}[hb!]
    \centering
    \includegraphics[width=0.8\columnwidth]{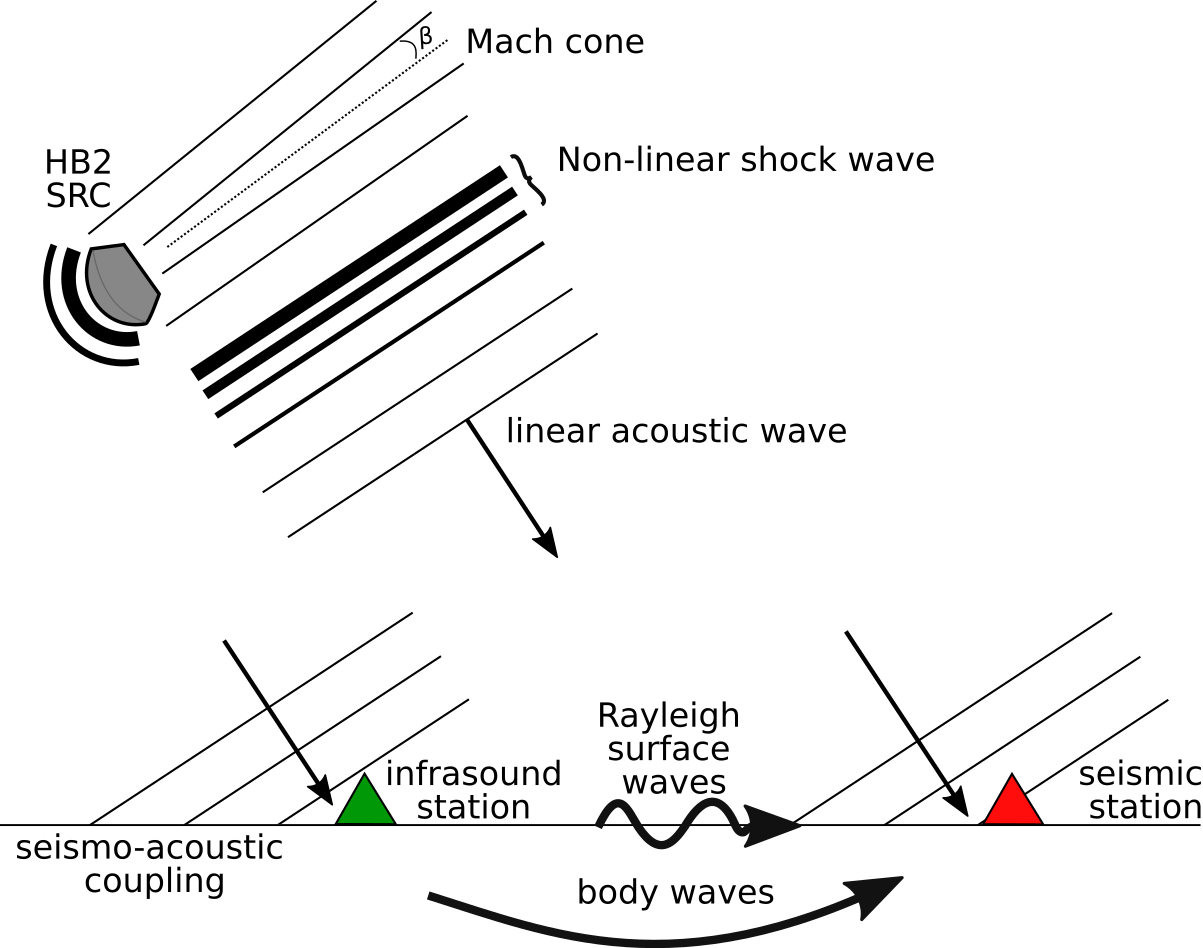}
    \caption{Schematic illustration of the shock wave generation during the hypersonic re-entry event of the Hayabusa-2 sample return capsule (HB2 SRC). Shock waves are generated by the Mach cone that travel almost perpendicular to the trajectory of the object, as the Mach cone angle $\beta$ --> 0 for such hypersonic trajectories. Non-linear shock waves rapidly decay to a linear wavefront that can be detected by infrasound sensors, as well as by seismic sensors. Air-to-ground coupling of acoustic waves can propagate as Rayleigh surface waves or body waves to seismic sensors. Figure redrawn from \citet{edwards2008seismic}.}
    \label{fig:shock_prop}
\end{figure}

From the hypersonic reentry of the Hayabusa-2 SRC in the upper and middle atmosphere, shock waves can be generated from the Mach cone along the SRC trajectory propagated through the atmosphere at a shallow angle (at about 12 degrees) with respect to the horizon (Fig. \ref{fig:shock_prop}). The angle of the Mach cone ($\beta$; Fig. \ref{fig:shock_prop}) can be calculated with respect to the Mach number of the reentry speed of the SRC (at about 12 km/s). For this event it will be approaching 0$^\circ$. The speed of sound at which linear acoustic waves travel is temperature dependent and is also affected by the upper and middle atmospheric wind profile.

Since the event of Hayabusa-1 re-entry, the team at Kochi University of Technology (KUT), collaborating with some manufacturing companies in Japan, have developed new infrasound sensors of INF01 and INF04 with integral data loggers. This allowed us this time to deploy significantly more systems that for the Hayabusa-1 campaign. 28 INF04 sensors were available to deploy in a 100 km scale target area of Woomera Prohibited Area (WPA).  

In addition to these systems, 7 small seismometers from Raspberry Shake (RS) were purchased in order to test these low-cost sensors and to confirm the air-to-ground coupling process. These consisted of   

\begin{itemize}
    \item 2x RS1D (1-Dimensional seismometer),  
    \item 2x RS3D (3-Dimensional seismometer), and 
    \item 3x RSB (Raspberry Shake \& Boom, 1-Dimensional seismometer and sonic-boom monitoring microphone) 
\end{itemize}

Note, see Table 1, and Raspberry Shake documentation\footnote{\footurlRS, \\ \footurlRSv} for specifications. 

There were also 2 sets of absolute nano-resolution barometers of Paro Scientific 6000-16B with a data logger (Mitomi Giken NL-6000), deployed with the INF04 sensors. The purpose of the absolute barometers was to calibrate the amplitude over pressure level at each site.  

Nationally available sensors for infrasound and seismic were also available. The global organisation of the CTBTO (Comprehensive nuclear-Test-Ban Treaty Organization) currently operates 52 infrasound arrayed observatories (currently of the total 60 planned sites worldwide), with 3 sites in operation in Australia. These are in Shannon (IS04), Hobart (IS05), and Warranmunga (IS07). The Australian National Seismograph Network (ANSN), operated by Geoscience Australia, also have several sites within the vicinity of the Hayabusa-1 SRC re-entry trajectory (See Fig. \ref{fig:main_map}). Some of these datasets may be used for confirmation of long-distant propagation possibility.

\subsection{Radio}
When a meteor enters the Earth’s atmosphere, it ionizes the atmosphere producing a characteristic plasma.  This plasma will strongly scatter in the VHF (30-300MHz) radio band, observable by the appropriate instrumentation \citep{maruyama_simultaneous_2006}. To detect this scattering, a VHF radio receiver can be located at a large distance from a transmitter at the same frequency, and this scatter observed. This system is known as radio meteor observation (RMO). A larger meteor, such as a fireball, directly emits plasma waves rather than just scattering reflections.   \citet{obenberger2014detection, obenberger2015dynamic, obenberger2016altitudinal} discovered that fireballs produce a radio afterglow at the HF (3-30 MHz) and VHF radio bands by the Long Wavelength Array (LWA1).
 
This direct plasma emission from the fireball is thought to be the same mechanism as solar flares, but it is not well understood. 
Type II or III solar radio burst are emitted within a few minutes of a solar flare and are related to shock waves and electron emission. This emission is considered to be a harmonic of the plasma frequency, but the conversion mechanism for the emission is still unknown. We think that the mechanism which the Hayabusa-2 SRC creates plasma and emits light is the same as that of a fireball. We therefore planned this observation of the plasma wave based on the idea that the same mechanism occurs as in type II or III bursts associated with solar flares. 
 
\citet{obenberger2016altitudinal} performed a statistical analysis of fireballs observed with the LWA1, 38 MHz radio telescope. At a luminosity of -4 Mag, the spectral flux density is 10$^4$ -- 10$^5$ Jy/s (Jy: Jansky). In this case, the typical meteoroid observed had a mass and entry velocity of 10 g and 30 km/s, respectively. The energies are therefore estimated to be 10$^7$ J using the average radio conversion efficiency of 0.1 - 1\%. The SRC is 16 kg, and will be entering at 12 km/s, resulting in 10$^9$ J of energy. Using the conversion efficiency predicted from the fireball, it is 10$^5$ -10$^7$ Jy. The minimum sensitivity of the instruments used is 10$^6$ Jy, which is sufficient for observation. 
 
The size of the meteoroids corresponding to the 10g of \citet{obenberger2016altitudinal} is about 1-5 cm, so the radio emission source is about 10 cm$^3$. The Hayabusa-2 SRC is however 40 cm in diameter, increasing this to 1000 cm$^3$, which is about 100 times larger. If the capsule is not compressed, the signal strength will increase 100 times, or if it is compressed, the density will increase. The density is proportional to the square root of the plasma frequency, and in this case the frequency increases by a factor of about 10. 
 
\citet{iwai2013peak} reported a spiky structure of less than 1 second in type-I bursts. Kaneda et al (2017) found a zebra pattern of about 1 second in type-IV. Therefore, we decided to record sub-second microstrucure by dividing 50-200 MHz by 200 and sweeping by $\frac{1}{4}$ second. 
 
\subsection{Energetic Particle Detector}
Two simple compact ionizing particle detectors were deployed at locations under the predicted trajectory \citep{aplin2017miniaturized}. These are sensitive to all ionising radiation capable of penetrating the instrument housing, including gamma rays and X-rays. This was somewhat of a speculative measurement, as no high energy particles have been reported from earlier spacecraft re-entry, although a few investigations have been published at energies beyond UV \citep{abe2011near, lohle2011airborne}. RaspberryPi computers were used as simple dataloggers for the detectors, powered by lead-acid batteries capable of operations for a few days without charging. 

\section{Experimental setup and Hayabusa-2 return scenario }
Hayabusa-2's sample return capsule was due to return on 5th December 2020 at 17:24 UTC. The predicted re-entry trajectory can be seen in Figure 3 and was approximately NW to SE. The fireball was predicted to have a peak brightness of -5 magnitude, entering the atmosphere at 12 km/s. 

\begin{figure*}[h!]
\begin{center}
    \includegraphics[width=\textwidth]{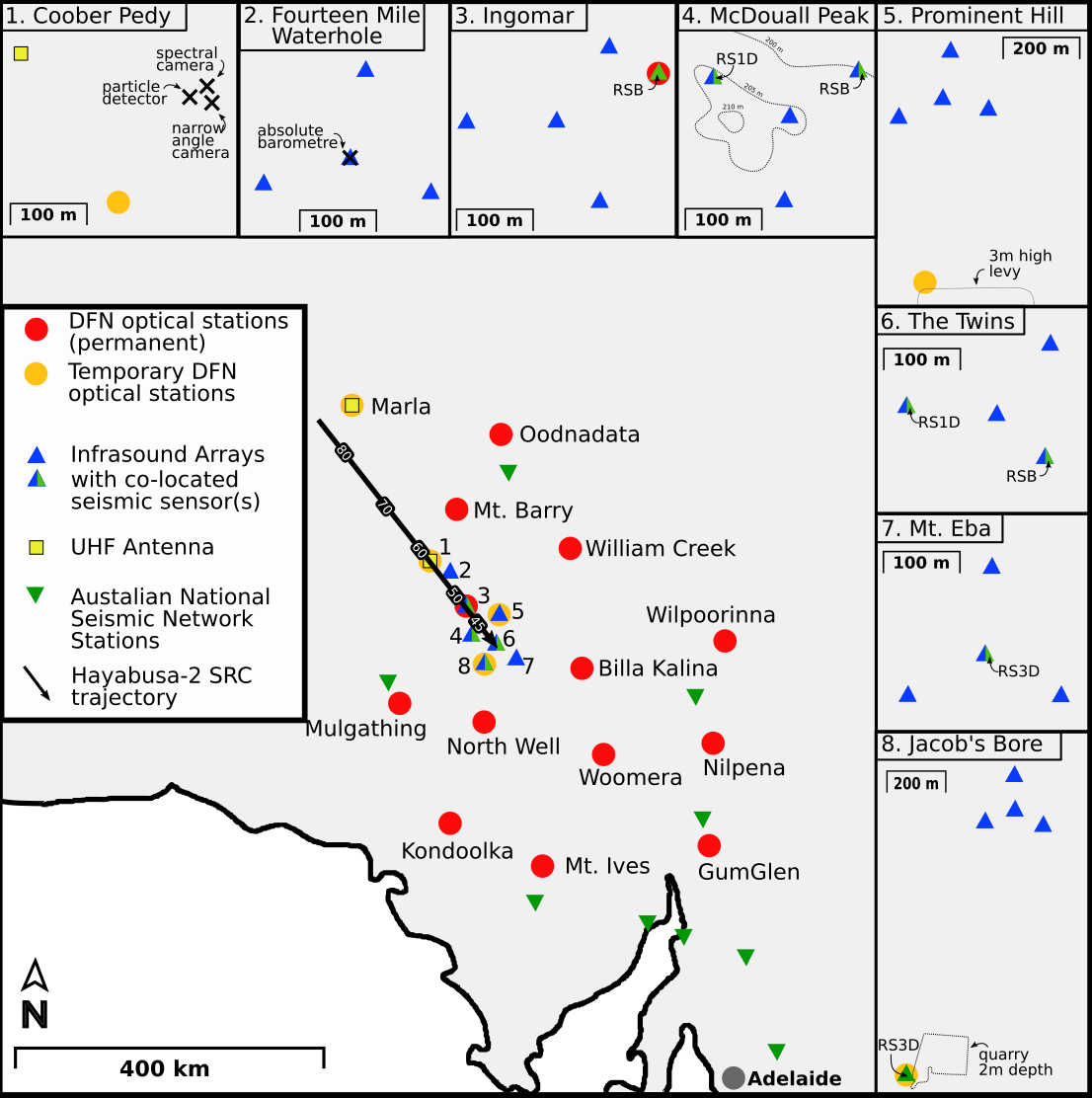}
    \caption{ Overview map of South Australia region showing the re-entry trajectory of the Hayabusa-2 sample return capsule. Approximate altitudes of the observed fireball are given along the trajectory arrow in km. This map illustrates the site locations for instruments deployed to observe the trajectory, including locations of permanent Desert Fireball Network and Australian Seismograph Network sites. The arrangement of infrasound sensors in each array at sites 2-8 are provided to scale, and oriented North up. }
    \label{fig:main_map}
\end{center}
\end{figure*}

Figure \ref{fig:main_map} shows the location of observation sites relative to the predicted re-entry of the Hayabusa-2 sample return capsule (SRC). For sites with multiple instruments, the relative set up positions are also shown to scale.

Due to COVID-19 constraints on travel, personnel on site was limited. In particular, entry restrictions into Australia prevented many scientists from attending. Volunteer support from interested amateurs was vital for systems deployment in the time available. However, personnel constraints meant that several instrumentation sites were deployed and activated prior to spacecraft arrival, but then had to be left unattended. Daily temperatures reached 46$^\circ$C, with night temperatures dropping as low as 13$^\circ$C. As well as contributing to restrictions of physical limitations on personnel, this significant variation may need to be taken into account for instrument records. 

Instruments at temporary sites were deployed 2 days prior to the re-entry event with a team of eight people. Marla, Coober Pedy, Prominent Hill and Jacob's Bore sites were attended on the night of the re-entry and these instruments were powered on around 12 hours before the event.

\subsection{Optical instruments}
Figure \ref{fig:main_map} shows the distribution of permanent Desert Fireball Network observatories in the area near the SRC re-entry. Four temporary stations were added close to the trajectory line for reasons described in Section \ref{sec:optic_instr}. Although the fish-eye lenses of the observatories enable capture of all-sky images, the sensor in the DSLR cameras crops a $5\degr$ section at the top and bottom. At permanent sites the crop sections are aligned North-South. For temporary sites, to maximise the capture of the fireball, the crop direction was aligned perpendicular to the trajectory at $\sim50\degr$. Permanent sites are powered by a solar power system, and temporary sites were powered by 140 Ah batteries.  

All-sky video were available at all temporary DFN sites, as well as Mt. Barry, William Creek, Ingomar and Billa Kalina. 

\begin{figure}[h]
    \centering
    \includegraphics[width=1.0\textwidth]{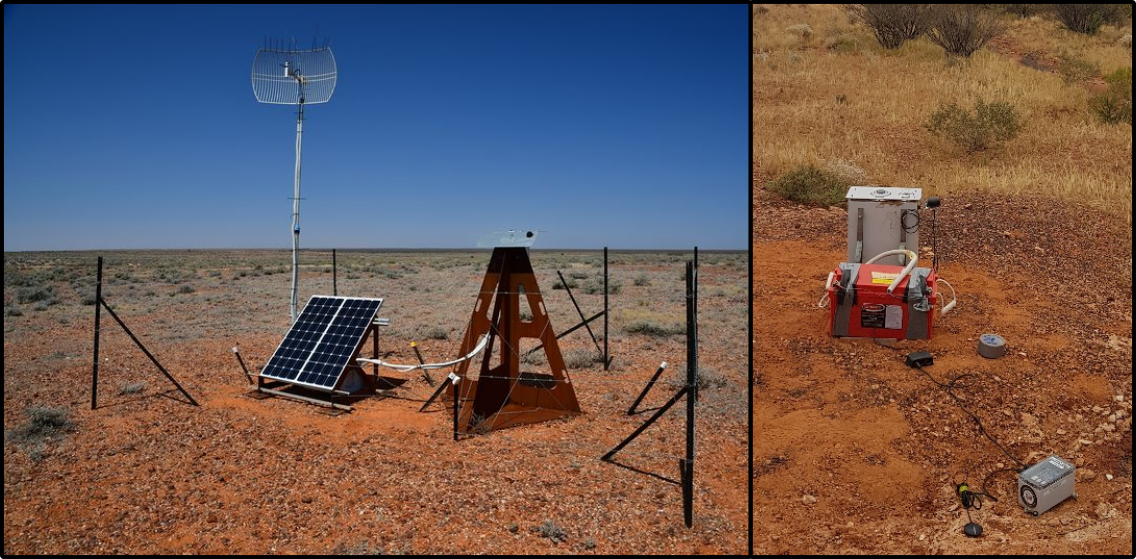}
    \caption{Left: Permanent DFN observatory site at Ingomar (Fig. \ref{fig:main_map}(site \#3)); Right:  Temporary DFN camera setup at Jacob’s Bore (Fig. \ref{fig:main_map} (site \#8), with 3D Raspberry Shake deployed on in situ bedrock.}
    \label{fig:dfn_cams}
\end{figure}

\begin{figure}[h]
    \centering
    \includegraphics[width=0.9\textwidth]{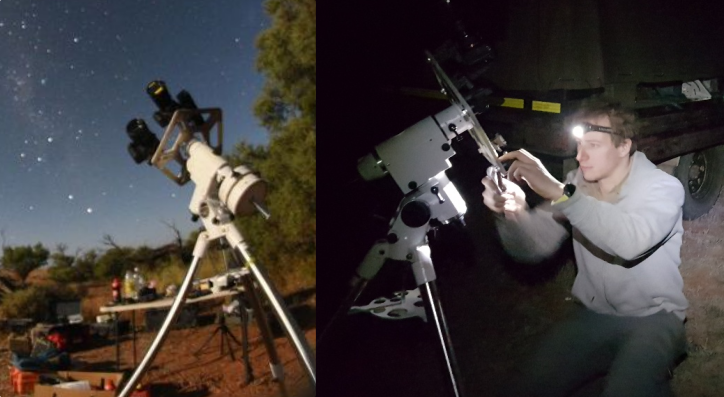}
    \caption{Set up of spectral and tracking video on a manually operated mount (not in final position).}
    \label{fig:tracking}
\end{figure}

At the Coober Pedy site, the fish eye spectral camera was set up, but due to power issues 10 minutes before the re-entry window, did not record. The narrow angle spectral camera was set up on an equatorial telescope mount with the ability to manually track the event (Fig. \ref{fig:tracking}).  

\subsection{Seismo-Acoustic instruments}
\begin{figure*}
\begin{center}
    \centering
    \includegraphics[width=1.\textwidth]{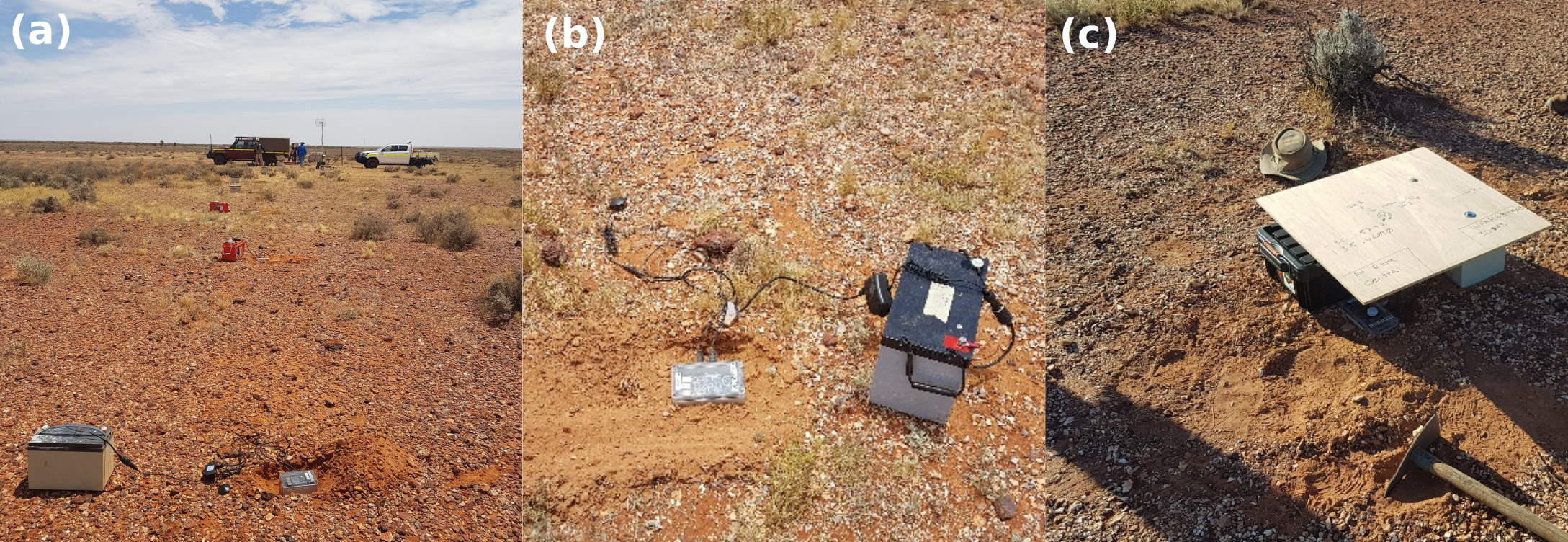}
    \caption{Testing of Raspberry Shake sensors (a,b). Sensors deployed in areas without bedrock were buried as seen in (c).}
    \label{fig:seismometers}
\end{center}
\end{figure*}

On the 1st Dec, a seismic line was set up using all 7 Raspberry shake sensors to test device timing and characterise a typical ground response. Sensors were buried in the hard-packed ground to increase ground coupling and to reduce wind noise that was significant that day (Fig. \ref{fig:seismometers}).  Test shots were taken with 5 m and 10 m spacing of instruments. 6 instruments correctly recorded data, with 3 showing correct timing from the USB GPS module. Software modifications were made to increase reliability of accessing GPS timing information and was tested on each system individually prior to SRC re-entry.  

\begin{figure*}
\begin{center}
    \includegraphics[width=\textwidth]{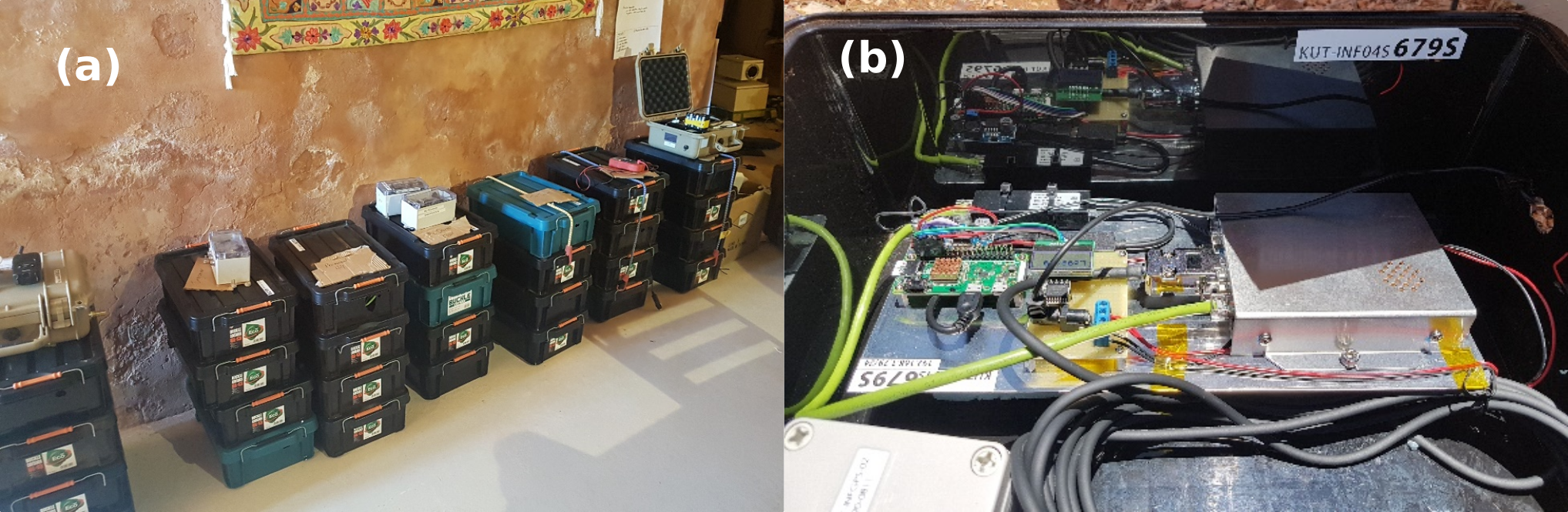}
    \caption{(a) seismo-acousic instruments ready for deployment on 3rd Dec., including infrasound systems. (b) interior setup of data logger, GPS and Raspberry Pi PC), Raspberry Shake instruments and absolute barometers.}
    \label{fig:infr}
\end{center}
\end{figure*}

\begin{figure*}
\begin{center}
    \includegraphics[width=\textwidth]{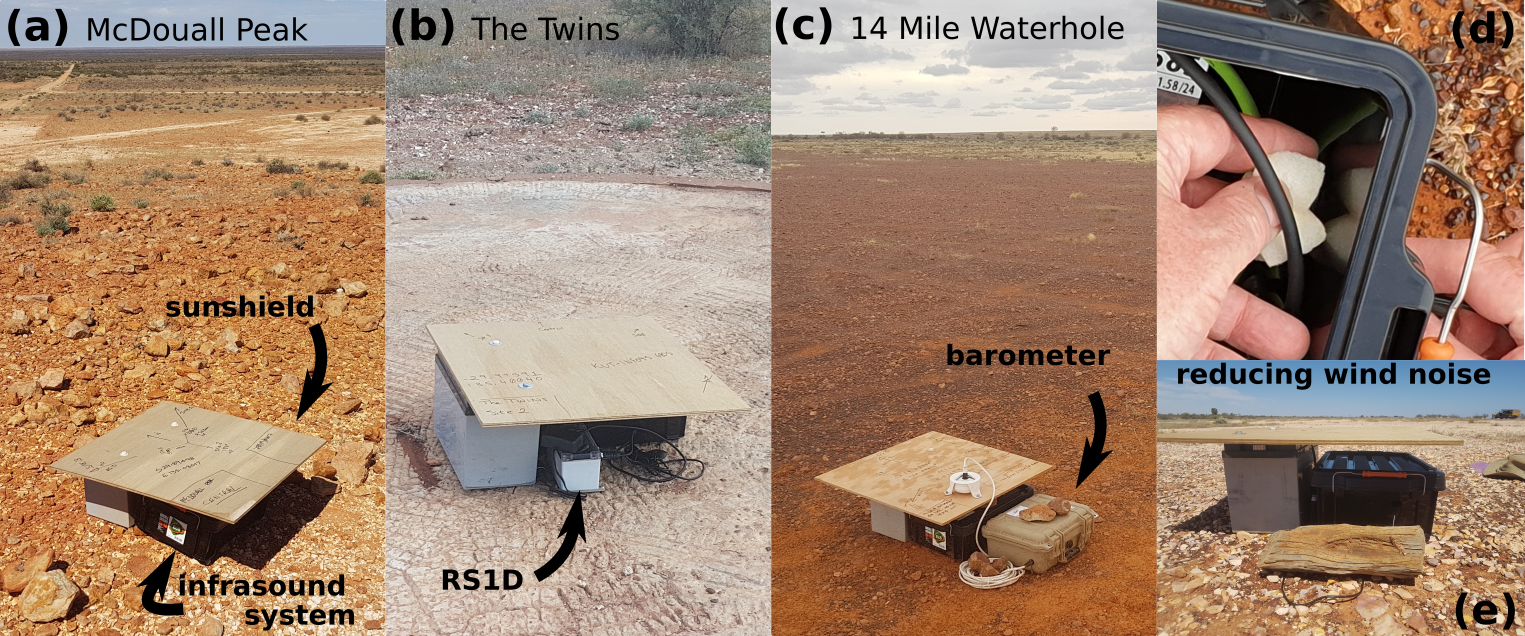}
    \caption{Deployment setup for seismo-acoustic sensors with 140 Ah battery supporting a plywood sunshield over black infrasound boxes. Raspberry Shake instruments can be seen in grey on the cement platform (b), and absolute barometer in beige pelican case in (c). Foam was used to decrease wind noise through power supply hole (d), and external cables were pinned to reduce vibration in the wind (e).}
    \label{fig:install_infr}
\end{center}
\end{figure*}

For detecting the precise trajectory and yield energy from the dataset of over-pressure amplitude at each site, we deployed 28 INF04 infrasound sensors at 7 sites on ground, with 4 sensors per array at each site (Fig. \ref{fig:infr}). At each seismo-acoustic site, infrasound sensors were installed with 140 Ah batteries capable of supplying the sensor for a few days. A sunshield was installed to protect the systems from the heat of the day, as they were installed 1-2 days prior to event entry (Fig. \ref{fig:install_infr}). Each sensor site was oriented from the central node using a handheld survey compass and distance measured using a 100 m tape to get relative alignment. Absolute position of the centre of the array was recorded using a handlheld GPS. Alignment of satellite nodes was intended to vary the acoustic arrival time, with close to 120$^\circ$ between each. Local topography and vegetation caused this to vary slightly across sites (Fig. \ref{fig:main_map}). Of the 28 sensors, some required rebooting on the day of the re-entry due to overheating. 

Seismometers were co-located with infrasound sensors at sites 3, 4, 6 and 8 (as per Fig. \ref{fig:main_map}).  

\subsection{Radio and Particle Detectors}

\begin{figure*}[h!]
\begin{center}
    \includegraphics[width=\textwidth]{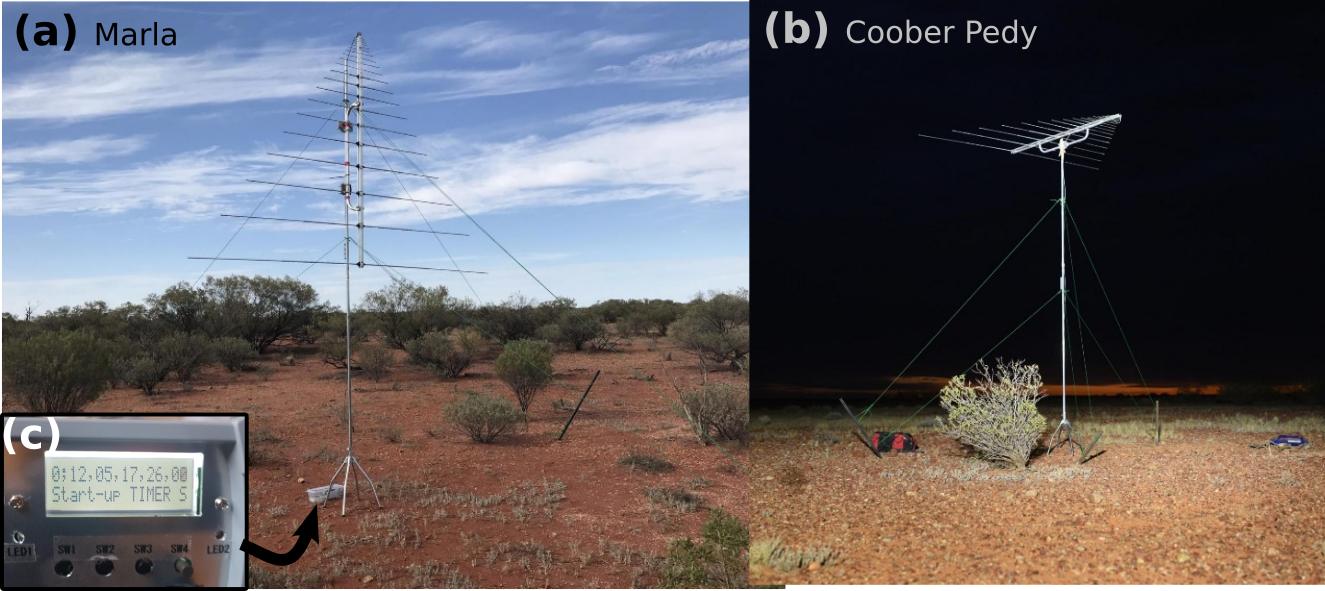}
    \caption{UHF setup in (a) Marla and (b) Coober Pedy. (a) is zenith aligned and (b) is horizontally aligned to an azimuth of 321$^\circ$. (c) shows programmed start time of receiver.}
    \label{fig:uhf_depl}

\end{center}
\end{figure*}

The UHF receiver is a modified version of e-callisto\footnote{\footurlCast}, and the antenna is used by UHF log-periodic antenna of the Japanese Creative Design Corporation. The receiver is displayed in Figure \ref{fig:uhf_depl}(c), the parts list and specifications in Table \ref{tab:uhf_parts} and \ref{tab:uhf_spec}. 
Two UHF antennas were installed; one up-range at the Marla site, the other at the Coober Pedy site (Fig. \ref{fig:main_map}). The Marla site was 46 km off of the predicted re-entry line to the East of the $\sim$90 km altitude point. The UHF antenna here was therefore set up pointing toward the zenith (Fig. \ref{fig:uhf_depl}a). The Coober Pedy site was almost directly below ($<$1 km East) the predicted 57 km re-entry point. The antenna was aligned horizontally toward an azimuth of 321$^\circ$ (Fig. \ref{fig:uhf_depl}b). The receivers for these antennas were programmed to begin recording at 2020-12-05T17:26:00 UTC. Energetic Particle detectors were co-located at these sites. 

\newpage
\section{Preliminary results with discussion}

\subsection{Optical instruments}
All DFN systems captured still images during the fireball event window and those with the capability to do so recorded monochrome video. The fireball from the Hayabusa-2 sample return capsule was recorded by nine DFN systems nearest the trajectory. Cloudy conditions across the region were partly responsible for obstructing the fireball low on the horizon for distant cameras. The additional narrow angle video camera at Marla also captured the fireball where it was below the sensitivity threshold for the DFN all-sky still and video at this site. Start times were staggered to allow continuous coverage of predicted re-entry times. The first visible point was seen from Marla at a height of 103 km at 2020-12-05T17:28:38.5 UTC, and the last visible point at 39 km from Billa Kalina at 2020-12-05T17:29:31.5. This is a total of 53 seconds of the SRC trajectory through the atmosphere. The maximum apparent magnitude seen by the video systems for the SRC was -5.1, based on a initial starfield calibration, in good agreement with the peak brightness prediction of -5.

\begin{figure*}
\begin{center}
    \includegraphics[width=1.\textwidth]{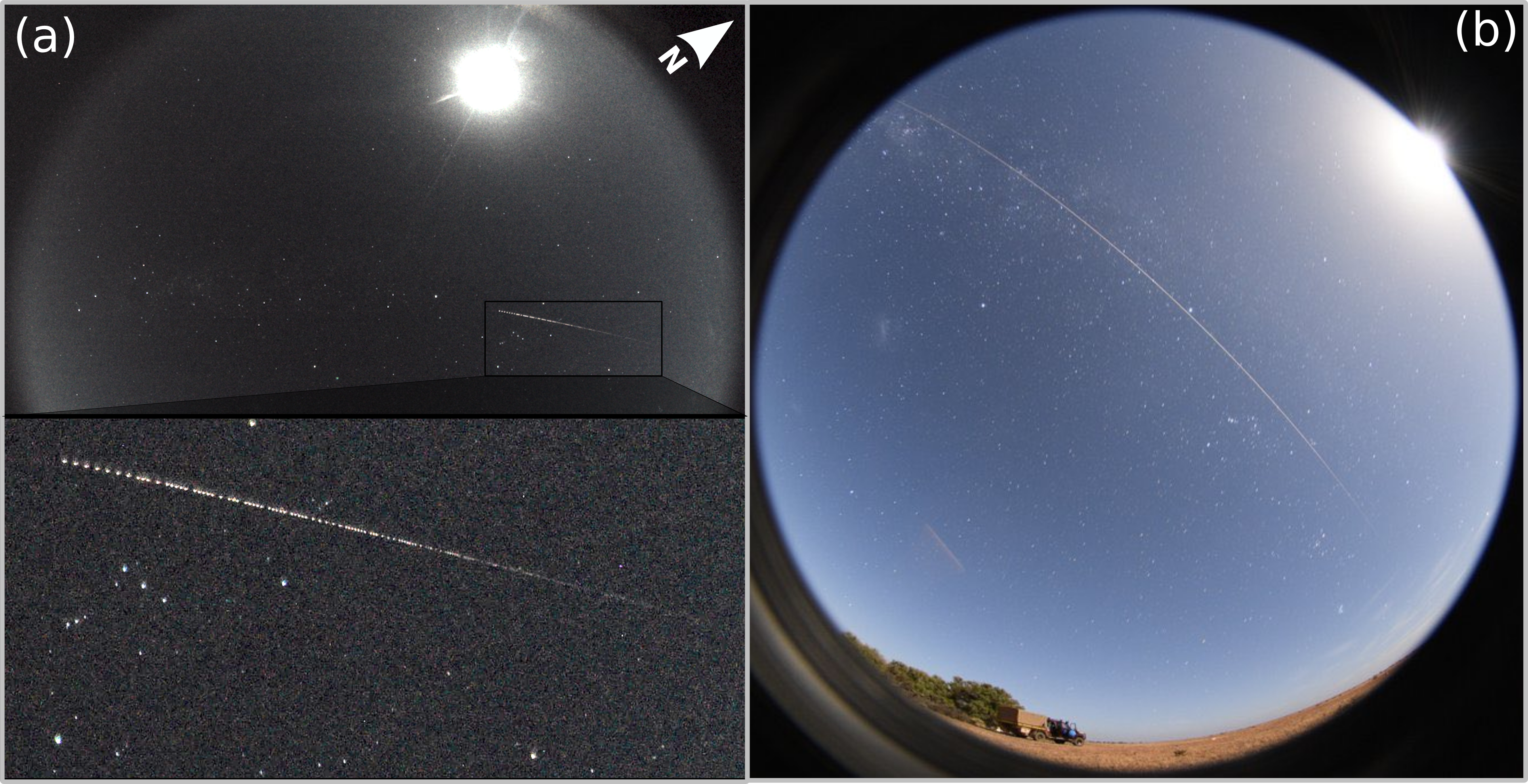}
    \caption{Image of the SRC fireball as seen from Coober Pedy using a DFN camera system (a; with zoom insert) and 'narrow angle camera' (Fig. \ref{fig:main_map} site \#1) (b). Note the appearance of a sporadic meteor in image (b), above the tree line.}
    \label{fig:dfn_images}
\end{center}
\end{figure*}

Using the new video systems was particularly advantageous for this event as the fireball created by the Hayabusa-2 SRC was a difficult target for the still imagers: 

\begin{itemize}
    \item it was a relatively faint object, therefore the extra sensitivity of the video was useful. 
    \item it had a slow apparent speed, leading to smudging of the shutter breaks in the still records (Fig. \ref{fig:dfn_images}). 
    \item the combination of Moon (73\% illuminated) and clouds at the time of entry significantly raised the sky background of the still images, further reducing their effective sensitivity. 
\end{itemize}

\begin{figure*}
\begin{center}
    
    \includegraphics[width=0.8\columnwidth]{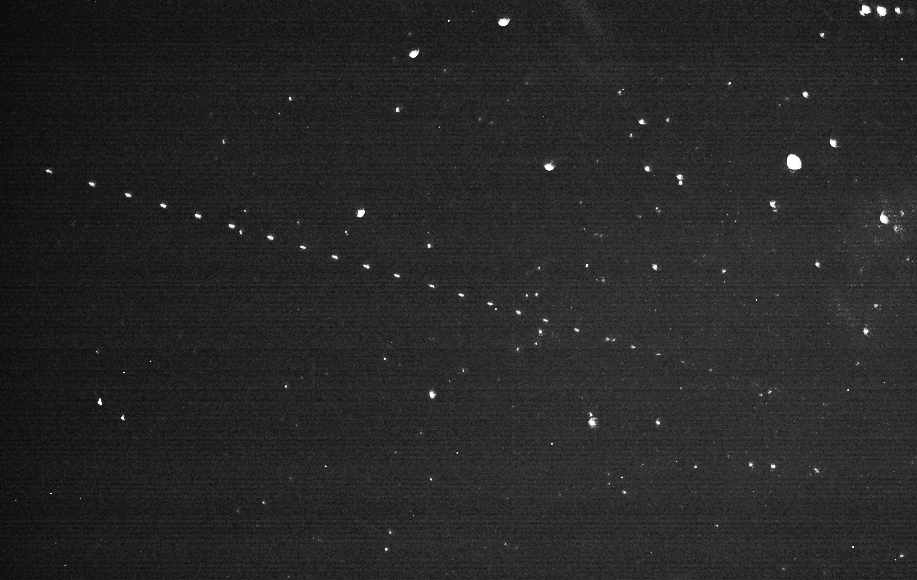}
    \caption{Stack of every 5th frame captured by the narrow video setup in Marla (Sec. \ref{sec:narrow_ang_vid}), showing the very beginning of the bright flight.}
    \label{fig:narrow_field_vid_stack}

\end{center}
\end{figure*}

\begin{figure}
    \includegraphics[width=0.8\columnwidth]{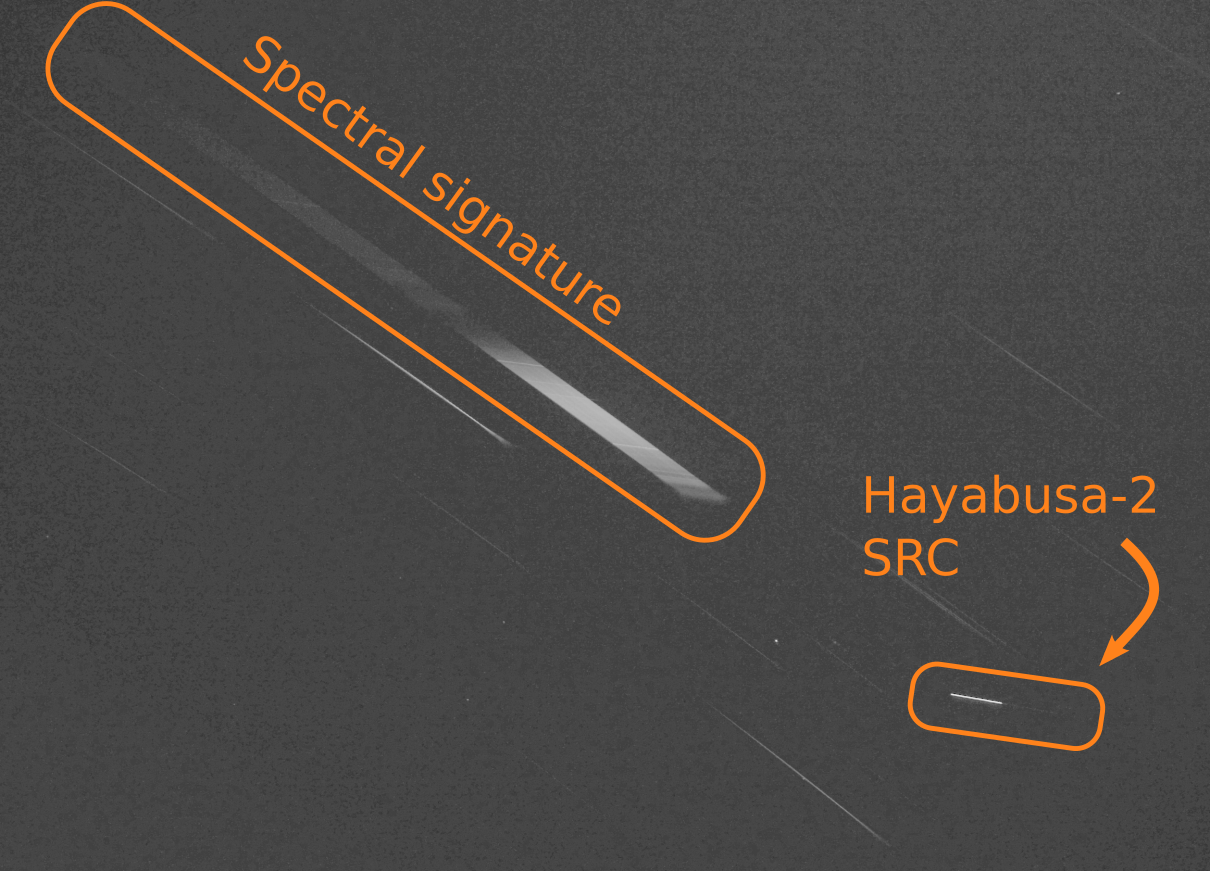}
    \caption{Single frame extracted from the spectral video.}
    \label{fig:spec_image}
\end{figure}

\begin{figure}%
    \centering
    \includegraphics[width=0.8\textwidth]{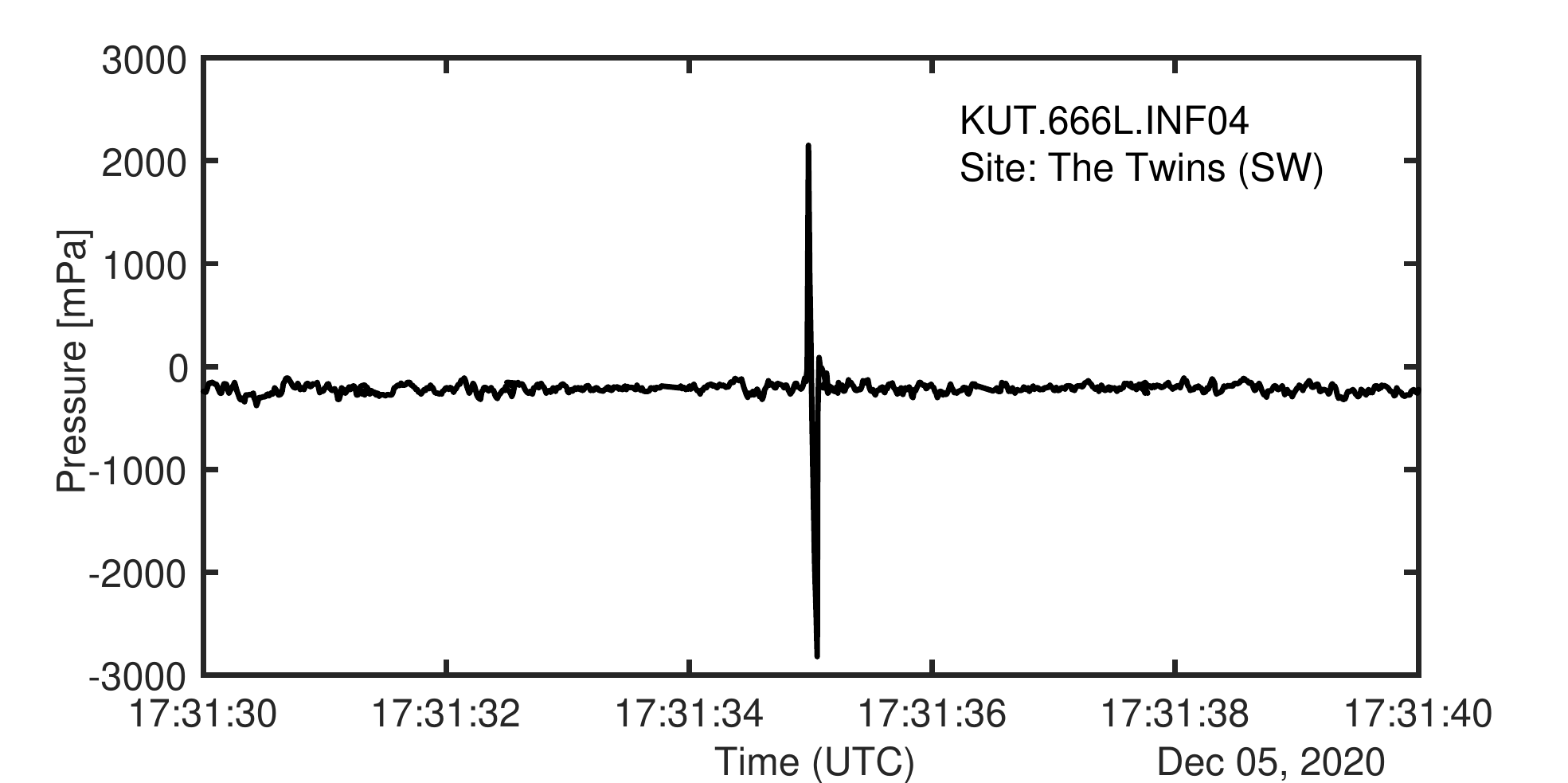}
    \caption{Example of the N-type infrasonic waveform recorded during the Hayabusa-2 SRC re-entry from sensor ID:666L at the arrayed site of The Twins. Note that the absolute value of the shown overpressure level is under calibration.}
    \label{fig:infra_wave}
\end{figure}%

The spectral video was collected by manually tracking the object on a pan tilt system. This was not ideal and significant shaking occurred, though video data were successfully recorded with the grating producing continuous spectra. Figure \ref{fig:spec_image} shows one of the video frames with the SRC and 1st and 2nd order spectra. The capsule spectrum is composed of gray-body and some emission lines in the near-ultraviolet region, which are molecular Nitrogen N2$^+$(1---) bands \citep{doi:10.1063/1.555546} from an atmospheric shock layer and CN violet bands from an ablating heat shield of a sample return capsule as seen in the former Hayabusa re-entry capsule \citep{abe2011near}. The wake emission of the SRC was also recorded which will be discussed in a forthcoming paper.

\begin{figure}%
    \centering
    \includegraphics[width=0.8\textwidth]{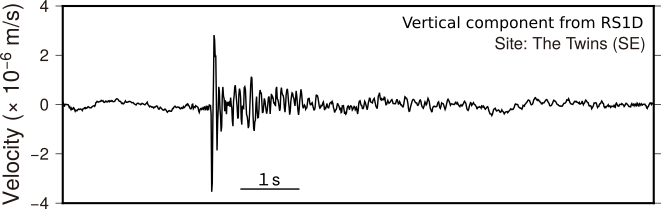}
    \caption{Example seismic waveform of the vertical ground velocity recorded during the Hayabusa-2 SRC re-entry at the Twins (SE site; see Fig. \ref{fig:main_map}, also pictured in Fig. \ref{fig:install_infr}(b).}
    \label{fig:seis_wave}
\end{figure}%
\subsection{Seismo-Acoustic instruments}
On hypersonic entry at 17:28 UTC  on Dec. 5, the shock waves from the Hayabusa-2 sample return capsule induced infrasound with a frequency range of about 0.5 Hz, just on the edge of the Mach cone (as per Fig. \ref{fig:shock_prop}). They travelled through the atmosphere to the ground at the speed of sound, propagating near-cylindrically in 3D space from the Mach cone. This passage through the atmosphere took over 3 minutes, and at about 17:31 UTC, the shock wave induced infrasound arrived at the first infrasound sensors. The arrival times propagated in order from the southmost site to the northmost one because of the smaller total distance from the SRC’s southbound trajectory to each site on the ground in 3D space, close to parallel to the normal vector from the Mach cone edge. 

Of the 28 infrasound sensors deployed,  the N-type waveform was successfully observed at 27 sensors. An example N-type waveform of the Hayabusa-2 SRC re-entry from the SW station of The Twins array is shown in Figure \ref{fig:infra_wave}. The N-type waveform was almost the same as for the Hayabusa-1 SRC reentry but without any complex features of the followed period by the fragmented Hayabusa-1 spacecraft. Of the 6 seismic sensors deployed, one sensor each at the Twins (Raspberry shake 1D), the McDouall Peak (Raspberry shake 1D), and the Mount Eba (Raspberry shake 3D) recorded successfully seismic waveforms excited by the induced shock waves from the Hayabusa-2 SRC (eg. Figure \ref{fig:seis_wave}). The peak ground velocities (PGV)  were comparable to those for the Hayabusa-1 SRC re-entry. The air-to-ground coupling process can also be investigated in comparison with previous result in 2010 \citep{ishihara2012infrasound}. A detailed investigation of the yield energy estimation and trajectory determination from this infrasound sensor network will follow in a future publication.  

Unfortunately the two Paro Scientific 6000-16B absolute nano-resolution barometers suffered power loss just short of the re-entry time.
This could be attributed to the cool desert temperatures at nighttime, compared to the expected battery life when testing in the lab prior to the event. 

Human ears can also hear the higher frequency audible range of the coming shock waves.
In 2010 for Hayabusa-1, an audio microphone detected such signals up to about 1 kHz at the impulsive signal of the SRC shock wave arrival \citep{yamamoto2011detection}. For this previous event, two observers also identified a large amplitude sound like thunder or fireworks. This time, ES and HD at the Coober Pedy site, noted a ``mine blast" like sound at 2020-12-05T18:32:15 UTC. This was a similar description to other members of the public in the Coober Pedy township. According to this reported time, the sound was not electrophonic in nature, but the conventional shock wave coming from the mach cone above the area with a delay of a few minutes from the witnessed artificial meteor.  The fact that the sound was noticed by more observers this time than in the Hayabusa-1 case is probably because the populated Coober Pedy site was almost directly under ground projection of the trajectory in the Hayabusa-2 case. Audio recordings were also made by the spectral video camera, although background noise levels were high, and processing will be required to confirm the audible signal arrival.

\subsection{Radio and EPD}
Unfortunately, no radio data were collected by the UHF antenna due to instrumental issues. The energetic particle detector at Marla also did not operate, but the detector at Coober Pedy successfully recorded data. Preliminary analysis shows no obvious increase in ionising particle counts compared to the background during the SRC re-entry, however we have yet to carry out a detailed statistical analysis to investigate the presence of any subtle features.

\section{Conclusions}
The Hayabusa-2's sample return capsule re-entered the Earth’s atmosphere over South Australia on the 5th December 2020 at 17:28 UTC. The hyper-sonic trajectory was visible as a fireball for over 53 seconds. The specific case of the Hayabusa-2 sample return capsule re-entry provides an example of the fortuitous use of existing fireball network (the Desert Fireball Network) to monitor and study a specific event, whereby the event can be used to calibrate the network, providing more insight onto existing astronomical data. A scientific observation campaign was planned to observe the optical, seismo-acoustic, radio and high energy particle phenomena associated with the entry of this interplanetary object. 49 instruments were deployed, with a further 26 permanent Desert Fireball Network sensors within range (total of 75). Observations were made of the Hayabusa-2 sample return capsule using the all-sky optical still and video cameras from the Desert Fireball Network, supported by additional optical narrower angle still and video systems, narrow angle spectral camera, and multiple arrays of seismic and infrasound sensors distributed around and along the reentry trajectory. Additional sensors deployed were two UHF antennae, and two energetic particle sensors.  Although some technical issues prevented full operations, 68 instruments successfully recorded data during the SRC arrival window, with positive detections of the phenomena on 38 of these. This is a high rate of success, and has acquired valuable data in optical and other non-optical measurements. Sufficient data have been collected to allow full trajectory reconstruction, and full analysis of seismic and infrasound data will give detailed insights into the energies generated from re-entering bodies. The Hayabusa capsule was observed during reentry from elevations of 103 to 39 km. Visual brightness peaked at an altitude of approximately 48 km at a magnitude of –5.1. The full duration of observed luminous reentry was 53.0 seconds, starting at 20-12-05T17:28:38.5 UTC.

\begin{table}[h!]
    \caption{Parts list of the UHF receiver.}
    \begin{tabular}{p{3.6cm} l}
\hline
\rowcolor[gray]{.95} RF Tuner & Philips Semiconductor \\[-5pt]
\rowcolor[gray]{.95}         & CD1316LS-I/V3\\[-5pt]
1Mbit SPI Serial & 23LC1024I/ST  \\[-5pt]
     SRAM & \\[-5pt]
\rowcolor[gray]{.95}1024k I2C & 24LC1025T-I/SN \\[-5pt]
\rowcolor[gray]{.95}     Serial EEPROM & \\[-5pt]
LOG Amp & AD8307ARZ  \\[-5pt]
\rowcolor[gray]{.95} MPU & PIC16F18456/SSOP  \\[-5pt]
Real-Time  & MCP7940N  \\[-5pt]
    Clock/Calendar & \\[-5pt]
\rowcolor[gray]{.95} GPS Module & GYSFFMANC  \\[-5pt]
SAW Components & B39871B3715U410  \\[-5pt]
\rowcolor[gray]{.95} LNA & MAAL-011139-TR1000 \\[-5pt]
Frequency Mixer & ADE-5 \\
\hline\\[-5pt]
    \end{tabular}
    \label{tab:uhf_parts}
\end{table}
\begin{table}[h!]
    \caption{Specifications of the UHF receiver.}
    \begin{tabular}{p{3.6cm} l}
 \hline
 \rowcolor[gray]{.95} Frequency range & 45 ~ 870 MHz  \\[-5pt]
 Frequency resolution & 100 kHz\\[-5pt]
 \rowcolor[gray]{.95} \multicolumn{2}{l}{Intermediate frequencies}      \\[-5pt]
\rowcolor[gray]{.95} \multicolumn{1}{r}{1st IF} & 869 MHz, \\[-5pt]
\rowcolor[gray]{.95}           & bandwidth 2 MHz\\[-5pt]
\rowcolor[gray]{.95} \multicolumn{1}{r}{2nd IF} &  10.7 MHz, \\[-5pt]
\rowcolor[gray]{.95}    & bandwidth 30KHz \\[-5pt]
\rowcolor[gray]{.95}     & and 300 kHz \\[-5pt]
 Dynamic range & -120 to -50 dBm\\[-5pt]
\rowcolor[gray]{.95}Noise figure & 4dB\\[-5pt]
 Channel sample rate & 800 channels/sec \\[-5pt]
      & (typically 200 channels\\[-15pt]   
      &  in 250 ms) \\[-5pt]
\rowcolor[gray]{.95}Frequency sweep rate & Less than 1 msec \\[-5pt]
 Time uncertainty & Less than 1 msec\\[-5pt]
\rowcolor[gray]{.95}Analog-digital & 14bit \\[-5pt]
\rowcolor[gray]{.95}    converter (ADC)  &\\[-5pt]
\rowcolor[gray]{.95}    resolution  &\\[-5pt]
 Interfaces & USB\\[-5pt]
\rowcolor[gray]{.95}RF  & input (N-F)\\[-5pt]
 Input voltage & 9 Vdc nominal \\[-5pt]
     & (6 ~ 15 Vdc) \\[-5pt]
\rowcolor[gray]{.95}Input current & 500mA \\[-5pt]
 Warm-up time & 1min \\[-5pt]
\rowcolor[gray]{.95} Weight & 0.52kg \\[-5pt]
 Dimensions & 130 x 115 x 70 mm\\[-5pt]
        & not including connectors \\[-5pt] \hline
    \end{tabular}
    \label{tab:uhf_spec}
\end{table}
\newpage
~\newpage
\section*{Acknowledgements}
Due to COVID-19 travel constraints both within Australia, and travel to and from Australia, the scientific team relied heavily on volunteers to help with logistics and operations of instrumentation. We are very grateful to Geoffrey Bonning, Andrew Cool, Karen Donaldson, Glyn Donaldson, Michael Roche and Terence Wardle; much of the observation campaign would not have been possible without their efforts. \\
The authors would like to express their sincere thanks to Dr. Nakazawa of
JAXA and the rest of the JAXA's Hayabusa-2 SRC recovery team.\\
The authors would also like to thank the support received from the Woomera Test Range, in particular Guy Mold and Colin Telfer; and the Defence Science and Technology Group, in particular Kruger White and team.  \\
This research is partly supported by The Murata Science Foundation and SECOM Science and Technology Foundation.\\
The Desert Fireball Network team is funded by the Australian Research Council under grant DP200102073 \\
EKS would like to acknowledge funding from the Institute of Geoscience Research and the Curtin faculty of Science and Engineering for travel and instrumentation funding.

\bibliographystyle{aasjournal}
\bibliography{biblio}

\end{document}